\documentclass{article}

\usepackage{arxiv}

\usepackage[utf8]{inputenc} 
\usepackage[T1]{fontenc}    
\usepackage{hyperref}       
\usepackage{url}            
\usepackage{booktabs}       
\usepackage{amsfonts}       
\usepackage{nicefrac}       
\usepackage{microtype}      
\usepackage{lipsum}\usepackage[utf8]{inputenc}
\usepackage{wrapfig}
\usepackage{float}
\usepackage{graphicx}
\usepackage{dcolumn}
\usepackage{url}
\title{A physical investigation on the phenomenon of the Dancing Coin}

\author{Barros, V. C.}

\author{Miguez, M. L.}
\date{\today}
\author{
  Victor C. C.~Barros \\
  Colégio Ari de Sá Cavalcante\\
  Fortaleza - Brazil\\
  \texttt{victorcortezcb@gmail.com} \\
   \And
 Malu M.~Miguez \\
  Colégio Farias Brito\\
  Fortaleza - Brazil \\
  \texttt{malumiguez@gmail.com} \\
}

\begin{document}
\maketitle

\begin{abstract}

When a coin is placed on the neck of a cold bottle, the coin will move cyclically. This is caused by the expansion of the gas inside the bottle warming up, which will cause the coin to rise when the internal pressure is large enough to overcome the weight of the coin. This problem, widely used to qualitatively demonstrate several laws of thermodynamics, has no quantitative solution in the literature. For this we use the ideal gas model and Newton's law of cooling combined with computational numerical simulations and systematic experimental measures to confront our theoretical model, including equations for the evolution of temperature, pressure and behaviour of the coins in the system.

\end{abstract}

\maketitle

\section{\label{sec:intro}Introduction}

The investigation of this phenomenon was first proposed as a problem for the International Physicists' Tournament 2018 \cite{iypt2018}, an international competition aimed at high-school students. The problem's statement was: 

\begin{quote}
``Take a strongly cooled bottle and put a coin on its neck.
Over time you will hear a noise and see movements of the
coin. Explain this phenomenon and investigate how the
relevant parameters affect the dance."     
\end{quote}
Published papers derived from IYPT problems are very common (\cite{eletromagcannon},\cite{iyptbook},\cite{iyptmagazine}). Due to that practice, the IYPT participants organized the ``IYPT Magazine", and many participants publish papers not only in the magazine, but in other scientific publications as well, one example is the book "International Young Physicists' Tournament: Problems \& Solutions 2012–2013" \cite{iyptbook}.
This paper is derived from the resolution developed by the authors to the problem in the competition, and may be adopted as both an educational resource in physics \cite{pedagogiciypt} as well as a scientific resource.

In this work, the authors use the ideal gas model \cite{idealgas}, alongside Newton's law of cooling \cite{newtonscooling} as the main basis to form a theoretical model of the phenomenon. Due to the discretization of some parts of the phenomenon (such as the coin jump), a computational solution to these equations was applied and compared to a completely analytical one. In order to validate the theoretical model, multiple experiments were performed, varying parameters such as bottle (shape and composition), coin weight, bottleneck area and temperature.

\section{\label{sec:theory}Theoretical Analysis}

\subsection{\label{sec:preliminary}Preliminary observations and qualitative analysis}

On preliminary observations of the phenomenon it was noted that the coin must be sealing the bottle in order for the phenomenon to occur. That is due to the agent causing the movement of the coin: the gas. It is clear that a leaking bottle will not be able to sustain an increase in pressure. The sealing can be made with the help of surface tension from water, which was shown to be effective, while applying negligible effects on the overall phenomenon. When the bottle is cooled the density of the gas (air) inside the bottle becomes higher, therefore when the bottle is removed from the colder ambient temperature and placed in the warmer ambient temperature, the density of the gas will reduce, as a result the pressure will increase since the number of moles of gas inside the bottle remains constant. This result can be explained by assuming that air is an ideal gas, eq. \ref{eq:idealgas}, and therefore follows the ideal gas law, which predicts an increase in $P$ (pressure) as $T$ (absolute temperature) increases while keeping $V$ (Volume) and $n$ (number of moles) constant.

\begin{equation}\label{eq:idealgas}
    P V = n R T
\end{equation}

When the pressure is sufficiently high so that the force applied on the coin due to the difference in pressure between the inside and the outside of the bottle is greater than the weight of the coin, the coin will lift and lose contact with the bottle neck, this causes the \textit{effective area} (the area in which internal pressure is applied to the coin) to shift from the bottleneck's area to that of the coin, which is always greater than the bottleneck's area, therefore the force applied on the coin will increase and further lift the coin. The process reaches its end as the gas loss reduces the pressure to ambient levels.

Once the coin falls back to the initial position and seals the bottleneck, a new cycle will start. At the beginning of each new cycle the outside of the bottle will be at ambient pressure and its internal temperature will be higher. Moreover, the number of moles of gas inside the bottle also decreases at each new cycle. These conditions result in an increasing interval between consecutive jumps.

\subsection{\label{sec:quant_analysis}Quantitative Analysis}

\subsubsection{\label{sec:level2}Coin Dynamics}
If the coin is approximately centred in the bottleneck (which is naturally a great portion of cases, due to the inner concavity of the bottleneck), the minimum pressure necessary to lift the coin can be calculated using the equilibrium of forces below:

\begin{equation}~\label{eq:eq_condition}
    \left(P_i - P_e\right) \cdot S = \Delta P \cdot S = Mg
\end{equation}

\noindent in which $P_i$ and $P_e$ are the internal and external pressures over the bottle, respectively, $S$ is the bottleneck's area, $M$ is the mass of the coin and $g$ is the local acceleration of gravity.

This is because when the coin is directly over the bottleneck, the area in equation \ref{eq:eq_condition} is the area of the bottleneck ($S$), but when the coin is raised the area is shifted from the bottleneck's area to the area of the coin (for a small lift), which is bigger, therefore increasing the upwards force on the coin. This means that when this minimum pressure is reached, the rest of the jump can be achieved, therefore it represents the minimum pressure difference needed to perform a jump.

\subsubsection{\label{sec:mechanisms}Heat Transfer Mechanisms}

We may consider three main mechanisms in which the bottle will exchange heat with the environment \cite{thermalbottle}, namely:

\subsection*{Conduction through the walls:}

As the width of the walls is much smaller than its surface dimensions, the heat flux through the bottle's walls $\phi_c$ can be approximated by Fourier's law:

\begin{equation}~\label{eq:conduction}
    \phi_c = \frac{dQ}{dt} = \frac{\kappa A \Delta T}{L} = C_{c} A \Delta T,
\end{equation}

\noindent in which $\kappa$ is thermal conductivity, $L$ is the width and $A$ is the surface area of the bottle walls. Here, $\Delta T$ represents the temperature difference between the air inside and outside the bottle. The heat flux due to conduction is expressed in terms of the coefficient $C_c$ which includes the properties of the wall.

\subsection*{Air convection}

Although air convection may not be analytically treated with high accuracy, a similar approximation can be used for small intervals of temperature \cite{heatprinciples}:

\begin{equation}~\label{eq:convection}
   \phi_v = \frac{dQ}{dt} =  (C_{e}  + C_{i}) A \Delta T
\end{equation}

\noindent in which the coefficients $C_i$ and $C_e$ represent air convection inside and outside the bottle, respectively.

\subsection*{Radiation}

If put in a decreasing scale, air convection is usually the most effective heat transfer way followed by conduction, and radiation heat transfer is the far less relevant heat transfer method\cite{thermalbottle}. However it is possible to be considered by using an approximation around equilibrium temperature:

\begin{equation}~\label{eq:radiation}
    \frac{dQ}{dt} =  C_{r} A \Delta T
\end{equation}

\noindent in which the coefficient $C_r$ includes the radiation heat flux properties around the equilibrium temperature of the system.

\subsubsection{\label{sec:heating}Air heating inside bottle}

Using equations \ref{eq:conduction}, \ref{eq:convection} and \ref{eq:radiation}, one can combine them into the Newton's law of cooling:

\begin{equation}
    \phi = \frac{dQ}{dt} =  k A \Delta T,
\end{equation}

\noindent in which $k$ is the cooling constant for the gas inside the bottle and the bottle itself. By direct comparison, it is possible to express the constant $k$ in terms of the heat transfer mechanisms as:

\begin{equation}
    k = \displaystyle \frac{1}{\displaystyle  \frac{1}{C_e+C_i}+\frac{1}{C_c}+\frac{1}{C_r}}
\end{equation}

The Biot number \cite{heatprinciples} of the system can be calculated as

\begin{equation}
    B_i = \frac{k L_c}{C_c} = \frac{k V}{C_c A}, 
\end{equation}

\noindent in which $L_c$ is the characteristic length of the bottle (usually defined as the ratio between the volume $V$ and the surface area $A$ of the Bottle, $V/A$). 

Through a calculation using preliminary data from the phenomenon, Biot numbers in the order of magnitude of $10^{-5}$ were found, in which case the bottle can be approximated as internally isothermal \cite{heatprinciples}.

An integration using the first law of thermodynamics arrives at a law for the heating of the bottle assuming a constant number $n$ of moles.

\begin{equation}
    \delta Q = dU + \delta W
\end{equation}

Assuming that in each jump we can consider $\delta W \approx 0$

\begin{equation}
    dU = (n C_v + C_g)dT = \phi dt = kA \Delta T dt
\end{equation}

Here, $\Delta T = T_a - T_g(t)$ represents the temperature difference  between ambient air and bottle air, respectively, and $C_g$ is the heat capacity of the glass.

\begin{equation}~\label{eq:temp}
    \frac{d T}{\Delta T} = \left(\frac{kA}{n C_v + C_g}\right)dt
\end{equation}

We can define a coefficient $b$ as:

\begin{equation}\label{eq:b_coef}
   b = \frac{kA}{n C_v + C_g}
\end{equation}

The $b$ coefficient includes the most important mechanisms involved in the cooling of the bottle, and it can be inferred through this calculation which parameters will affect it. However, it must be measured experimentally for each individual bottle since the convection coefficients are very difficult to measure. It can be also estimated computationally using Computational Fluid Dynamics and Heat Transfer models \cite{usageofcfd}.

Using the coefficient $b$ the time dependence of the temperature can be expressed as:

\begin{equation}
\label{eq:temperature}
   T_g(t) = T_a - (T_a-T_0)\cdot e^{-bt}
\end{equation}

\noindent since $T_0$ is the initial bottle temperature. By combining this expression and the ideal gas law to obtain the time dependence of the pressure inside the bottle during the interval between the jumps, the pressure can be shown to be

\begin{equation}~\label{eq:pressure}
    P = \frac{nR[T_a - (T_a-T_0) \cdot e^{-bt}]}{V}
\end{equation}

 It is necessary to discretize equation \ref{eq:pressure} since when the coin jumps the pressure will become the same as ambient pressure. Therefore the expression is only valid as long as the bottle is sealed from ambient pressure. Another important remark is that the equalisation with ambient pressure is due to the lost moles of gas, which will not be considered in the analytical calculations.

\subsubsection{\label{sec:time_jumps}Time interval between jumps}

Another important factor to be analysed in the phenomenon is the time it takes for the coin to jump, or the time interval between consecutive jumps, which, as said in the qualitative analysis, tends to increase with time. One can arrive at the analytical expression for this time using the same approximations used before.

As a fixed difference of pressure $\Delta P$ is needed to lift the coin and that the outside pressure remains constant, the expression for the difference in temperature needed for one jump, assuming constant number of moles as an approximation, can be derived. Here is important to recall that as the time interval the bottle remains unsealed the amount of gas which escapes is small in comparison to the total amount of gas inside the bottle, which justifies our constant number of moles assumption. The expression we obtain using the ideal gas law is

\begin{equation}
\label{eq:tempvar}
   \Delta T_i = \frac{V \Delta P}{n R}
\end{equation}

Which is the temperature difference for the $i$-th jump as a function of the pressure difference expressed in equation \ref{eq:eq_condition}. As a result the temperature between the jumps is constant. Using equation \ref{eq:b_coef} for the $\Delta T_i$ temperature:

\begin{equation}
   \Delta T_i = T_{i}-T_{i-1}
\end{equation}

\begin{equation}
   T_i - T_0 = \sum^{i}_{n=1} \Delta T_n
\end{equation}

\begin{equation}~\label{eq:temperatura_inicial}
   \Delta T_i  = (T_i-T_0) - \sum^{i-1}_{n=1} \Delta T_n \\
               = (T_i-T_0) - \frac{(i-1) V \Delta P}{nR}
\end{equation}

Then, substituting $T_i$ from equation \ref{eq:temperature}, but using it to calculate the time needed $\Delta t$ for the temperature to change by $\Delta T_i$ the equation for the interval of time it takes the coin to jump based on the index $i$ of the jump can be found. Using $\Delta T_f = T_a - T_0$

\begin{equation}
    T_i = T_a - (T_a - T_{i-1})\cdot e^{-b \Delta t}
\end{equation}





\begin{equation}
   \Delta t(i) = \frac{\ln \left(1+ \frac{\Delta T_i}{\Delta T_f - i \Delta T _i}\right)}{b}
\end{equation}

\subsubsection*{Comparison between analytical and numerical method}

\begin{figure}[H]
    \centering
    \includegraphics[width=8cm]{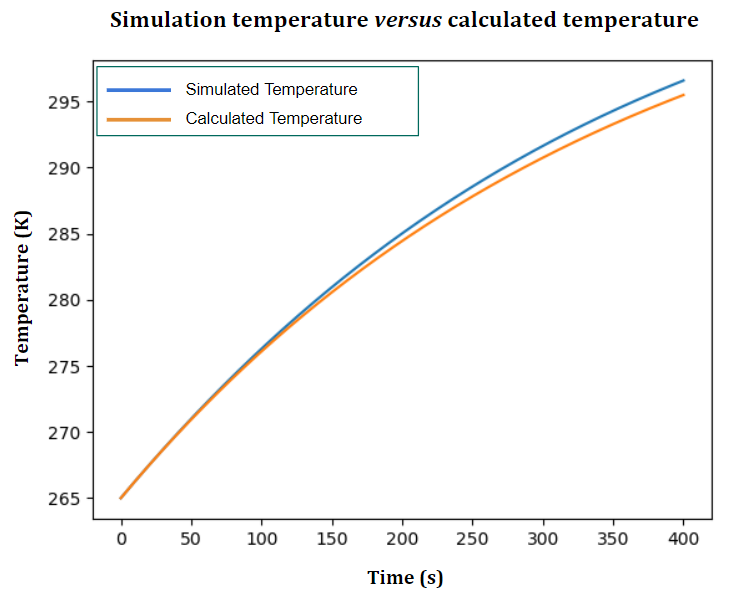}
    \caption{Comparison between the time dependence of the temperature for both the numerical Python simulation (in blue) and the analytical model \ref{eq:temperature} (in orange). The simulation takes into account the variable number of moles in each discrete jump. For long periods of observation, the discrepancy of the results with both methods is within the acceptable limits of the experiments.}
    \label{fig:comp_num_vs_analy}
\end{figure}

As a method to account for the variation of moles and the discretization of the pressure due to the discrete jumps, a computer numerical solution of the equations \ref{eq:temp} and \ref{eq:tempvar} was proposed using Python, in order to achieve more accurate results, as it can be seen in figure \ref{fig:comp_num_vs_analy}.

\begin{figure}[H]
    \centering
    \includegraphics[width=9cm]{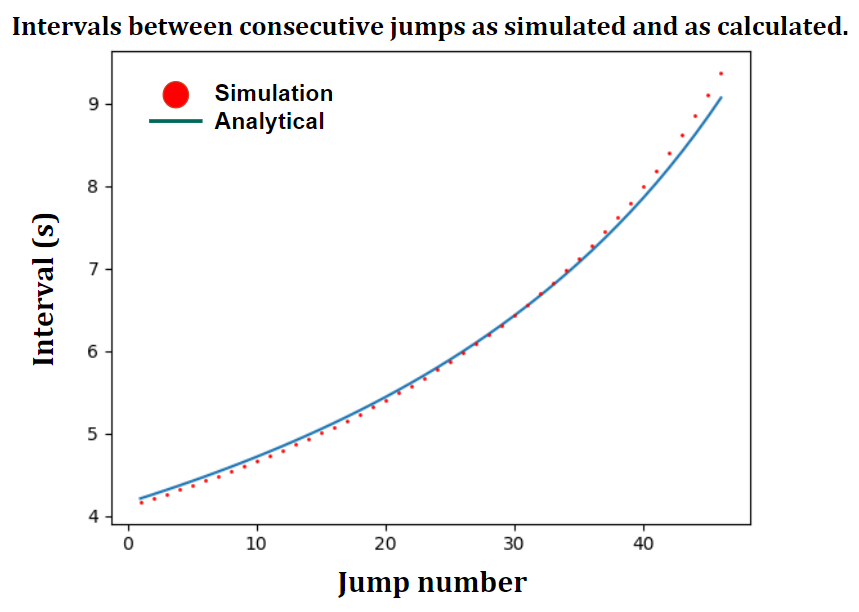}
    \caption{Comparison between the numerical calculation for the interval between jumps accounting for the discretizations and the analytical calculations.}
    \label{fig:time_interval}
\end{figure}

The difference between the two calculations starts to be relevant for time $t > 400$s, which is long enough when compared with the typical time interval between the jumps in our experiment, as can be seen in figure \ref{fig:time_interval}.

\begin{figure}[H]
    \centering
    \includegraphics[width=8.5cm]{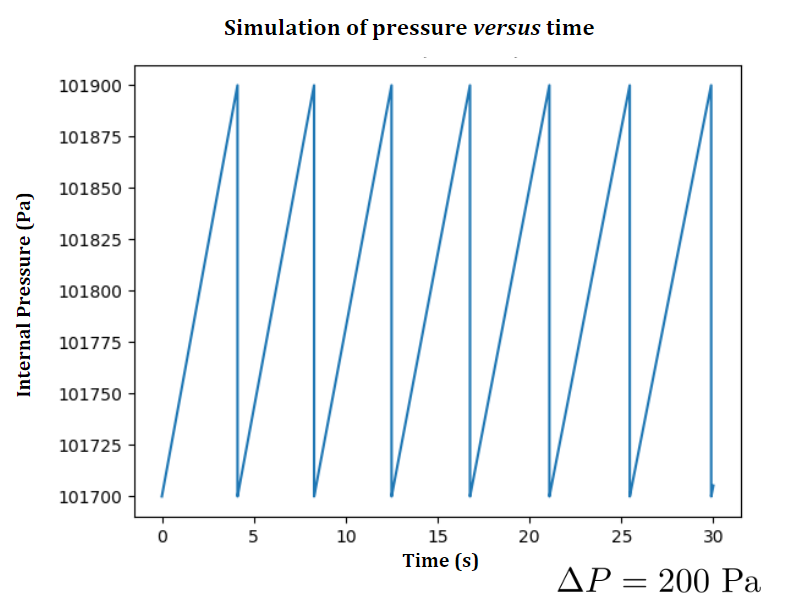}
    \caption{Small variations in the pressure inside the bottle calculated numerically. In the figure we take into account the discrete jumps and moles lost and $\Delta P$ fixed at $200$Pa.}
    \label{fig:pressure}
\end{figure}

In figure \ref{fig:pressure} it is possible to observe the cyclical nature of the pressure inside the bottle. It is noticeable that the relative variations in pressure are very small, around $0,2\%$. Since the time interval between the jumps is very small it may seem that the pressure rises linearly, however it varies exponentially between jumps.

\section{\label{sec:exp_investigation}Experimental Investigation}

\subsection{\label{sec:parameters}Relevant parameters for investigation}

The medium and temperature in which the bottle is placed are extremely relevant, since the difference in temperature is the main driving force for the jumps. On the subject of the dynamics of the coin, we will analyse the effects of the area of the bottle neck and the mass of the coin in order to validate the theoretical predictions. On the subject of the thermodynamics of the bottle we will analyse how different bottles transfer heat at different rates.

\subsection{\label{sec:materials}Materials and Method}

\begin{figure}[H]
    \centering
    \includegraphics[width=9cm]{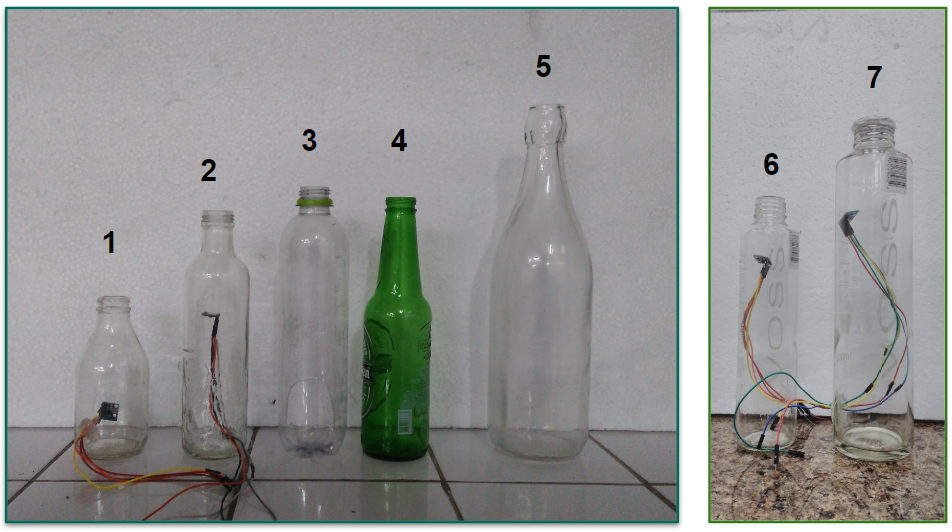}
    \caption{Bottles used in the experiments. From left to right: 
1 - Small Bottle (Glass)
2 - Medium Bottle (Glass)
3 - Plastic Bottle (Plastic)
4 - Reference Bottle (Glass)
5 - Big Bottle (Glass)
6 - Small Cylindrical Bottle (Glass)
7 - Big Cylindrical Bottle (Glass)
}
    \label{fig:bottles}
\end{figure}

We have used seven different bottles (see figure \ref{fig:bottles}), six of which were made of glass, and one made of plastic, which was included in order to analyse the effects of the material composing the bottle. The bottles have different shapes and sizes to allow a good qualitative explanation of the differences in each of the bottle's $b$ coefficient, but two of the bottles are cylindrically shaped and scaled sizes of one another, allowing for the variation of the scaled size alone.

For the coins, we used current Brazilian Real coins, alongside little optional bolt washers to be included on top of the coins in order to vary the weight of them, and also additional coins varying from a quarter dollar coin to old Brazilian coins. We expect that as long as the coin is sufficiently smooth so that it can seal the bottle appropriately, no significant difference should be expected from the shape of the coin alone, but rather from the mass of the coin, which is the main parameter in the coin.

\begin{figure}[H]
    \centering
    \includegraphics[width=0.3\textwidth]{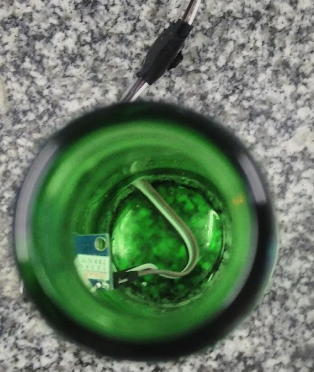}
    \caption{BMP180 sensor used to obtain temperature and pressure data of air inside the bottle.}
    \label{fig:sensor}
\end{figure}

The equipment chosen to measure temperature and pressure inside the bottle was the BMP180 \cite{bmp180}, which was used because it is a tiny sensor widely used nowadays (see figure \ref{fig:sensor}. Since the volume of the sensor is negligible when compared to the volume of the bottle, we expect that it shall cause no disturbance in the measurements whatsoever. Data was obtained at a sampling rate of 10 ms.


The sensor can measure temperature at a precision of $\pm 0.1$ K and pressure at a precision of $\pm 1$ Pa \cite{bmp180}, which is sufficient for the measurements taken, and has the critical advantage of being very low-cost and small enough to fit inside the bottle and not interfere significantly with the experiment. The sensor was used in a dedicated module board, wired through a small hole made in the base of the bottle, the hole was posteriorly sealed with hot glue. The wires were connected to an Arduino Uno board \cite{arduino} which sends the data to a computer in real time through the serial port. The data acquired in the computer is further analysed. The cooling method used was a freezer part of a common household fridge. It is important to remark that we only need to ensure that the bottle has a low enough temperature for the phenomenon to be observed. By inspection we observed that a high precision control of the initial temperature (around $273$ K) of the bottle was not necessary for an ambient temperature around $298$K. Moreover, if the observation of a specific interval of temperature is desired, it is sufficient to allow the bottle to cool below the interval, and wait for it to heat until it reaches the desired interval.

In order to observe the jumps of the coin, the cold bottle was placed on a plane table and the coin was put over its neck using water to seal the system. After that, it was ensured that the bottle was not affected by any external disturbance.


\section{\label{sec:Exp_results}Experimental Results}
\subsection{\label{sec:P_behaviour}Overall Pressure Behaviour}

In order to analyse the effects of coin's mass, surface area of the bottleneck, initial bottle temperature as well as size, format and composition of the bottle, we performed a series of experiments for which the results we present in the following sections.

Experiments were made, varying mass of the coin, area of the bottleneck, temperature and bottle, including the two geometrically equivalent bottles and bottles with very different materials and wall thickness.

\begin{figure}[H]
    \centering
    \includegraphics[width=8cm]{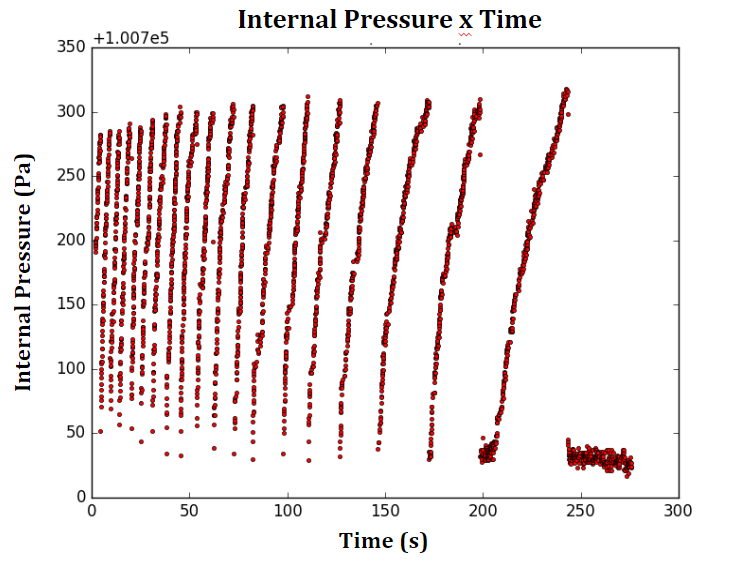}
    \caption{Air pressure inside the bottle (number 4 shown in figure \ref{fig:bottles}) as a function of time. The steps represent the jumps of the coin. Experimental data from pressure as a function of time inside bottle number 4.}
    \label{fig:pressure_steps}
\end{figure}

In figure \ref{fig:pressure_steps} it is shown the air pressure inside the bottle as a function of time for bottle number 4 (shown in figure \ref{fig:bottles}) when completely sealed. The steps of pressure shown in figure are related to the coin jumps. In the experiment it is clearly seen that the pressure difference between successively steps is almost the same, as predicted by our theoretical assumptions.

In the first experimental analysis, it is clearly noticeable that the pressure indeed follows the cyclical nature predicted earlier. Showing a consistent pattern in all experiments in which the bottle did not leak.

\subsection{\label{sec:time_interval}Time intervals for consecutive jumps}

\begin{figure}[H]
    \centering
    \includegraphics[width=8cm]{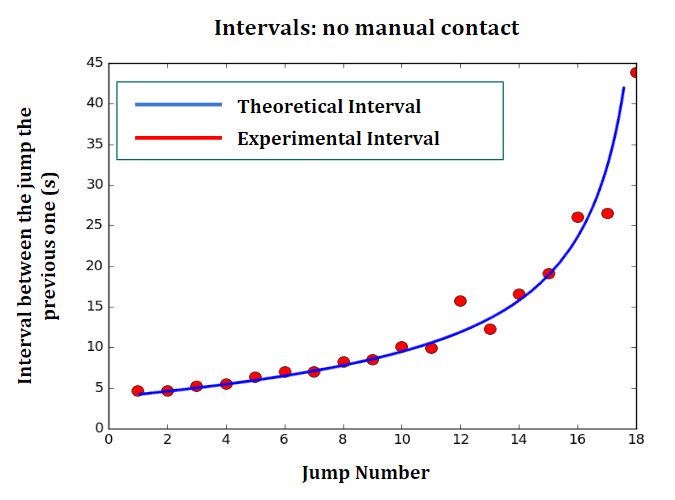}
    \caption{Comparison of typical experimental data for the time interval between consecutive jumps for bottle number 1, shown in figure \ref{fig:bottles}, to theoretical prediction from numerical simulation.}
    \label{fig:time_steps}
\end{figure}

Using the peaks found in graphs as shown in figure \ref{fig:pressure_steps}, one can obtain the time interval between peaks, therefore the time interval between coin jumps. Although in our assumptions we assumed that there were no leakage during the experiments, the comparison between experimental results with our theory are in a good accordance, as it can be seen in figure \ref{fig:time_steps}.

\begin{figure}[H]
    \centering
    \includegraphics[width=8cm]{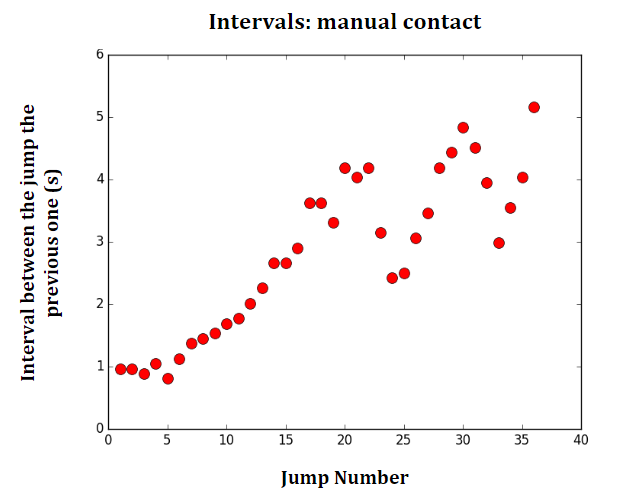}
    \caption{Experimental data for the time interval between consecutive jumps for bottle number 4 (shown in figure \ref{fig:bottles}) when it is hold by hand.}
    \label{fig:holding_bottle}
\end{figure}

As a matter of additional investigation, the effects of holding the bottle was made. In figure \ref{fig:holding_bottle} we present the time interval between jumps of the coin when we hold the bottle by hand. It is observed that after some time the time interval does not increases as predicted by our theory.

With manual contact, no simple mathematical law was found due to the irregularity of the flux. Along with more irregular data, the main result was that with manual contact the bottle will heat much faster. A possible explanation is the intensification of convection.

\subsection{\label{sec:temperature}Temperature}

\begin{figure}[H]
    \centering
    \includegraphics[width=8cm]{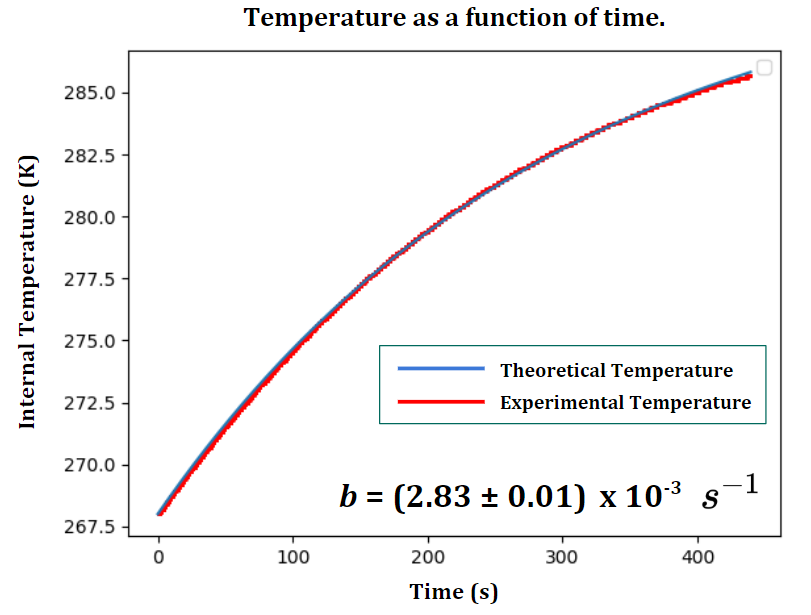}
    \caption{Experimental data for temperature as a function of time for bottle number 4 (shown in figure \ref{fig:bottles}) compared to the theoretical prediction.}
    \label{fig:temperature}
\end{figure}

In figure \ref{fig:temperature} we show the internal temperature of the bottle as a function of time. As it can be compared to our theoretical model, Newton's law of cooling correctly describes the behaviour of the temperature dependence, and reinforces that the heat exchanges during coin jumps can also be neglected. To observe this behaviour the room temperature, in which the experiment was performed, was maintained constant during the observed phenomenon.

The agreement is usually exceptionally good, with correlation coefficient $r$ above $99\%$ in most cases. For the results to be precise the room must remain in constant temperature, and that was achieved through careful control of the refrigeration of the room in which the measurements were taken.

\subsection{\label{sec:area}Area of the bottleneck}

Using the five non-cylindrical bottles and measuring the pressure difference needed to raise the same coin it is possible to verify the dependence on area calculated before \ref{eq:eq_condition}, with good agreement. The way to measure the pressure difference at a jump is to simply measure the pressure difference between the peak and the following valley at the pressure plot.

\begin{figure}[H]
    \centering
    \includegraphics[width=9cm]{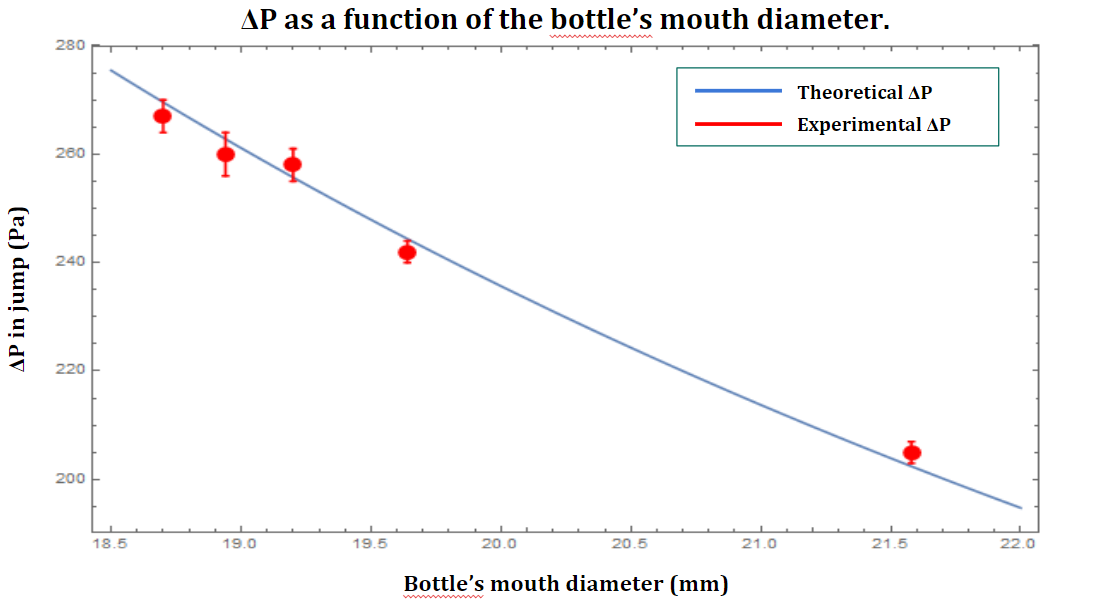}
    \caption{Points from multiple bottles relating the theoretical and experimental diameter of the bottleneck with the pressure difference needed to raise it.}
    \label{fig:pdif_diameter}
\end{figure}

As it can be seen in figure \ref{fig:pdif_diameter}, the good agreement between experiment and theory for different bottles shows that their shape does not affect the dependence of the observed phenomenon.

\subsection{\label{sec:coin_mass}Mass of the coin}

\begin{figure}[H]
    \centering
    \includegraphics[width=8cm]{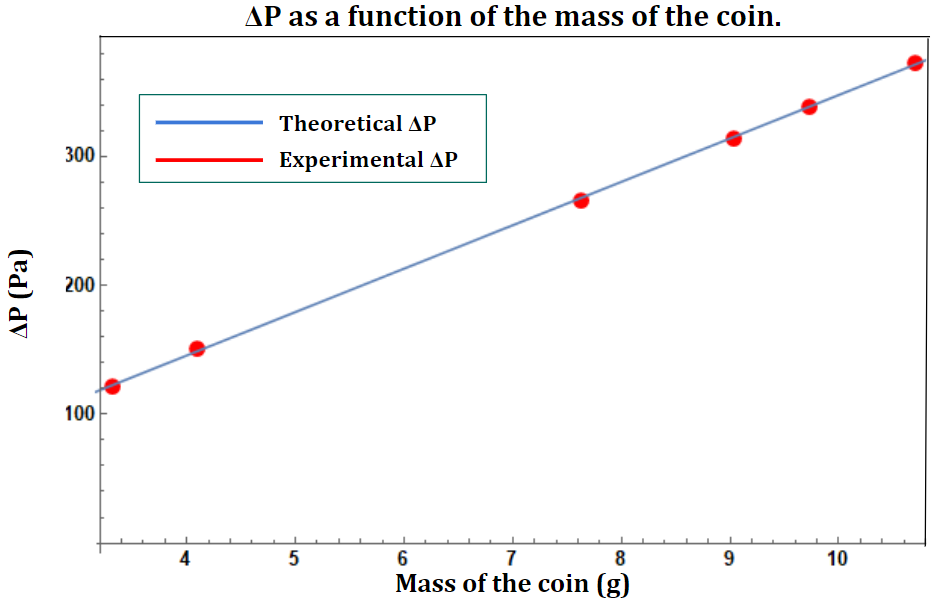}
    \caption{Comparison of the experimental average pressure difference necessary to lift each coin with the theoretical calculation. The average was taken for at least 50 jumps for each coin.}
    \label{fig:coin_mass}
\end{figure}

We used the coins and the additional weights to vary the weight and using the same procedure as subsection \ref{sec:area}, the average value for $\Delta P$, over multiple jumps, could be calculated for each coin weight. These results are shown in figure \ref{fig:coin_mass}.

\subsection{\label{sec:bottle}The bottle}

Using a computational fit for equation \ref{eq:temperature}  on the data of temperature as a function of time, the value for $b$ coefficient for each bottle can be calculated. Results are shown in table \ref{tab:b_coefficient}.

\begin{table}[H]
 \centering
 \begin{tabular}{||c|c||} 
  \hline
 Bottle & $b (s^{-1}\cdot 10^{-3})$ \\ [1ex] 
 \hline\hline
 1 & $4.50 \pm 0.02$  \\ 
 \hline
 2 & $3.70 \pm 0.02$ \\
 \hline
 3 & $10.3 \pm 0.4$\\
 \hline
 4 & $2.83 \pm 0.01$\\
 \hline
 5 & $2.63 \pm 0.01$\\
 \hline
 6 & $3.42 \pm 0.01$\\
 \hline
 7 & $2.77 \pm 0.07$ \\ [1ex] 
 \hline
\end{tabular}
\caption{Average value for $b$ coefficient as measured for each bottle.}
\label{tab:b_coefficient}
\end{table}

Those measurements agree with our qualitative analysis. For example, bottle 3 is made of a thin plastic wall, therefore its conduction heat coefficient is high, which can be seen by its $b$ coefficient, since it is one order of magnitude above all the others. We also observe that bottle 7, which is a bigger version of bottle 6, presents a lower $b$ coefficient, which also agrees with qualitative analysis since heat conduction is inversely proportional to bottle size.

\section{\label{sec:sound}Sound of the jumps}

Since the phenomenon is popularly known as the ``Dancing Coin'', it was also investigated the sound emitted by the coin during the jumps. When the coin leaves or reaches the bottleneck it emit a sonic pulse. Each of these pulses were recorded, and using the software Audacity \cite{audacity} the sound was finely cleaned as to isolate the sound of the phenomenon. The average frequency was obtained by a Fourier transform of the signal recorded by selecting the peak frequency.

\begin{figure}[H]
    \centering
    \includegraphics[width=9cm]{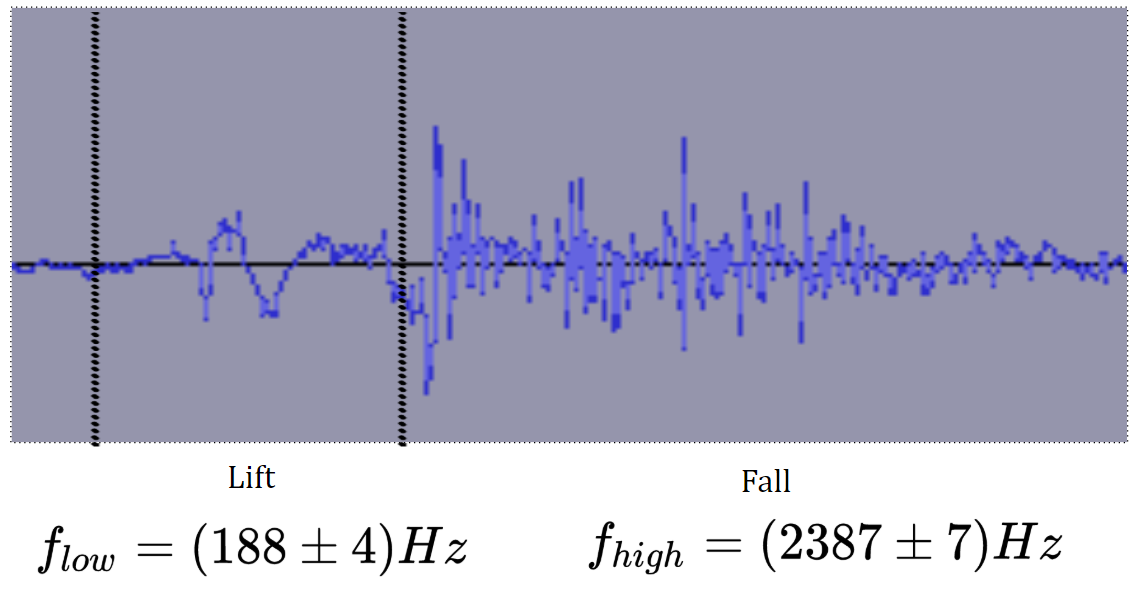}
    \caption{Graphical visualisation of the sound of one jump in bottle 4. Lines were traced to divide between the two distinct regimes.}
    \label{fig:sound}
\end{figure}

The sound is constituted of two main parts, the lifting part, in which the coin is losing contact with the bottle, in this part a deep sound is heard, which can be easily explained qualitatively as due to the bottle becoming a sound tube and the escaping gas causing that effect. When the coin hits the bottleneck the sound heard is sharper, which can also be explained as the characteristical sound of metal hitting glass, which is known to be sharp.\cite{metalsound}

\section{\label{sec:conclusion}Conclusion}

In this work we proposed a model for the behaviour of the ``dancing coin", and using an extensive experimental analysis of several parameters, such as time evolution of pressure and temperature of air inside the coin, we compared with good agreement a set of analytical equations, and a numerical solution to a set of experimental data. The experiments have shown the cyclical nature of the phenomenon as well as how the time scale of these cycles evolve with time as was predicted b our theory. We also analysed the so called \textit{sound of the jumps}, which was isolated as a result of two parts of the phenomenon, the rising and falling of the coin, and were shown to present a characteristic frequency for each part.


\section{\label{sec:appendix1}Appendix}
\subsection{\label{sec:vander}Comparison with Van der Waals' model}
It may be considered that the ideal gas model is not adequate for this system, but its difference from the ideal gas model when taking into account difference in pressure, is small enough to be less significant than experimental errors. Van der Waals equation:

\begin{equation}
    \left(P+\frac{n^2 a}{V^2}\right) \left(V-nb \right) = nRT
\end{equation}

Variation in pressure with Van der Waals model:

\begin{equation}
    \Delta P_v = \frac{nR(T_1-T_2)}{V-nb}
\end{equation}

Variation in pressure in the ideal gas model:

\begin{equation}
    \Delta P_i = \frac{nR(T_1-T_2)}{V}
\end{equation}

Using the values for a common bottle and air: $n = 0.0147\,mol$, $b = 0.0387 \frac{L}{mol}$, $V = 0.33L$, a simple relation between the two calculations can be achieved by:

\begin{equation}
  \frac{\Delta P_i}{\Delta P_v} =  \frac{V-nb}{V}\approx 0.998
\end{equation}

\noindent Which is an error of 0.2\%. This simple comparison shows that the error is minimal for the systems investigated in this work.

\bibliographystyle{unsrt}  
\bibliography{references.bib}  

\begin{thebibliography}{10}

\bibitem{iypt2018}
John Balcombe, Samuel Byland, and Ilya Martchenko.
\newblock Problems for the 31st iypt 2018, Jul 2017.

\bibitem{eletromagcannon}
Saba Zargham and Hamid Ghaednia.
\newblock No. 1, electromagnetic cannon: Measurements and simulations of the
  terminal velocity of a projectile.
\newblock {\em IYPT Proceedings Book 2010—2011}, 1(1), Jan 2012.

\bibitem{iyptbook}
Sihui Wang and Wenli Gao.
\newblock {\em International Young Physicists' Tournament}.
\newblock {WORLD} {SCIENTIFIC}, sep 2014.

\bibitem{iyptmagazine}
Iypt-magazine.
\newblock http://iyptmag.phy.ntnu.edu.tw/.

\bibitem{pedagogiciypt}
Gorazd Planinsic.
\newblock Iypt problems as an efficient source of ideas for first-year project
  laboratory tasks.
\newblock {\em European Journal of Physics}, 30(6):S133, 2009.

\bibitem{idealgas}
Kevin~M Tenny and Jeffrey~S Cooper.
\newblock {\em Ideal Gas Behavior}.
\newblock StatPearls Publishing, 2018.

\bibitem{newtonscooling}
R.~H.~S. Winterton.
\newblock Newtons law of cooling.
\newblock {\em Contemporary Physics}, 40(3):205–212, 1999.

\bibitem{thermalbottle}
Robert~T. Bailey and Wayne~L. Elban.
\newblock Thermal performance of aluminum and glass beer bottles.
\newblock {\em Heat Transfer Engineering}, 29(7):643--650, jul 2008.

\bibitem{heatprinciples}
Frank~P. Incropera.
\newblock {\em Principles of heat and mass transfer}.
\newblock John Wiley, 2013.

\bibitem{usageofcfd}
Zhiqiang Zhai and Qingyan~(Yan) Chen.
\newblock Numerical determination and treatment of convective heat transfer
  coefficient in the coupled building energy and cfd simulation.
\newblock {\em Building and Environment}, 39(8):1001–1009, 2004.

\bibitem{bmp180}
Bosch.
\newblock {\em BMP180 Data sheet}, 5 2015.
\newblock 0 273 300 244.

\bibitem{arduino}
Arduino website.
\newblock https://store.arduino.cc/usa/arduino-uno-rev3.

\bibitem{audacity}
Audacity website.
\newblock https://www.audacityteam.org/.

\bibitem{metalsound}
Roberta~L. Klatzky, Dinesh~K. Pai, and Eric~P. Krotkov.
\newblock Perception of material from contact sounds.
\newblock {\em Presence: Teleoperators and Virtual Environments},
  9(4):399–410, 2000.

\end{thebibliography}

\end{document}